\documentclass[aps,prd,preprint,showpacs,nofootinbib]{revtex4}
\usepackage{graphicx}
\usepackage{color}

\begin{document}
\title{ Transverse momentum resummation in soft-collinear effective theory}
\author{\center{Yang Gao, Chong Sheng Li\footnote{\hspace{-0.1cm}Electronics address: csli@pku.edu.cn},
and Jian Jun Liu \footnote{\hspace{-0.1cm}Present address:
Department of Physics, Tsinghua University, Beijing 100084,
China}}} \affiliation{\small Department of Physics, Peking
University, Beijing 100871, China}

\begin{abstract}
We present a universal formalism for transverse momentum
resummation in the view of soft-collinear effective theory (SCET),
and establish the relation between our SCET formula and the well
known Collins-Soper-Sterman's pQCD formula at the next-to-leading
logarithmic order (NLLO). We also briefly discuss the
reformulation of joint resummation in SCET.
\end{abstract}

 \maketitle

\section{Introduction}

Recently, the soft-collinear effective theory (SCET) has made
great simplifications on the proof of factorization in B meson
decays \cite{bsr} and high energy hard scattering processes
\cite{ee,hsf}, including resummation of large logarithms in
certain regions of phase space, for example, $e^{+}e^{-}$
annihilation into two jets of thrust $T\rightarrow1$ \cite{ee},
the deep inelastic scattering (DIS) in the threshold region
$x\rightarrow1$ \cite{dis} and Drell-Yan (DY) process in the case
of $z\rightarrow1$ \cite{dy}. The reason for these facts is that
SCET can be viewed as an operator realization of the pQCD analysis
when the modes participating the interactions of interest are soft
and collinear, just like chiral dynamics vs QCD at low energy
region. This effective field theory (EFT) provides a simple and
systematic method for factorization of hard, collinear and usoft
or soft degrees of freedom at operator level, especially usoft
modes can be decoupled from collinear modes in the Lagrangian at
leading order by making a field redefinition, and the large double
logarithms such as
$(\alpha_s\log^{2}\frac{Q^2}{\Lambda^{2}})^{n}$, where $Q,\Lambda$
are two typical scales that characterize a process, can be
resummed naturally through the running of renormalization group
equation (RGE).

However, all the above works have not discussed the transverse
momentum ($Q_T$) distributions of high energy hard scattering
processes. In this paper, we will investigate the resummed $Q_T$
distributions \cite{abc}, taking the Higgs-boson production via
gluon fusion in small $Q_T$ region \cite{kt,pt} as an example,
within the framework of SCET. It can be seen that in SCET the
$Q_T$ resummation formula automatically separates the
process-dependent Wilson coefficient and universal anomalous
dimension of the effective operator in a process, which once has
been studied by the authors of \cite{upt} within the
Collins-Soper-Sterman (CSS) frame.

The paper is organized as follows. In section II we start by
reviewing the basic steps for factorization and resummation in
SCET. In section III we apply it to derive the $Q_T$ distribution
at small $Q_T$ region directly, which confirms the CSS formula. In
section IV, we also discuss a similar formula for joint
resummation. Section V contains our concluding remarks. The
details of calculation are given in Appendix.

\section{Preliminaries}

SCET is appropriate for the kinematic regions of collinear and
usoft (soft) modes with momenta scaling:
$p_{c}=(p^{+},p^{-},p_{\perp})=(n\cdot p,\bar{n}\cdot
p,p_{\perp})\sim Q(\lambda^{2},1,\lambda)$ and $p_{us}\sim
Q(\lambda^{2},\lambda^{2},\lambda^{2})$ or $p_{s}\sim
Q(\lambda,\lambda,\lambda)$, where $\lambda \ll 1$ is the scaling
parameter, and the light-like vectors $n=(1,0,0,1)$,
$\bar{n}=(1,0,0,-1)$ satisfy $n\cdot\bar{n}=2$, and the
perpendicular components of any four vector $V$ are defined by
$V_{\perp}^{\mu}=V^{\mu}-(n\cdot
V){\bar{n}^{\mu}}/{2}-(\bar{n}\cdot V){n^{\mu}}/{2}$.

In constructing SCET, one should first identify the scaling of all
possible modes of initial and final states with soft and collinear
degrees of freedom, then integrate other degrees of freedom, and
the remaining modes must reproduce all the infrared physics of the
full theory in the region where SCET is valid, which is ensured by
the method of regions for Feynman integrals with massless quarks
and gluons\footnote{In the presence of masses, the regions
analysis is very complicated, and we only discuss the massless
case. } \cite{sr}. The EFT describing the usoft(soft) and
collinear modes is known as SCET$_{\mathrm{I}(\mathrm{II})}$, and
to distinguish the two theories, the scaling parameter
corresponding to SCET$_{\mathrm{II}}$ is denoted by $\eta\sim
\lambda^{2}$, i.e., $p_{c}\sim Q(\eta^{2},1,\eta)$ and $p_{s}\sim
Q(\eta,\eta,\eta)$.

The elements of SCET$_{\mathrm{I}}$ consist of usoft sectors
\{$q_{us},A_{us}$\} and collinear sectors \{$\xi_{n},A_{n}$\}
moving in the $n$-direction, which are expanded as
\begin{eqnarray}
\phi_{n}(x)&=&\sum_{\tilde{p}}e^{-i\tilde{p}\cdot
x}\phi_{n,p}(x),\qquad\ p=\tilde{p}+k,
\end{eqnarray}
where $k\sim Q\lambda^{2}$ resides in the space-time dependence of
$\phi_{n,p}(x)$, i.e., $\partial\phi_{n,p}(x)\sim
(Q\lambda^{2})\phi_{n,p}(x)$, and $\tilde{p}\sim Q(0,1,\lambda)$
is called label momentum, and the label operators
$\bar\mathcal{{P}}$, $\mathcal{P}_{\perp}$ are defined by picking
out $\tilde{p}^{-}$, $\tilde{p}_{\perp}$ momenta for collinear
fields $\phi_{n}(x)$\footnote{The convention
$\phi_{n}(x)=\phi_{n,p}(x)$ for collinear fields will be used for
convenience.}, respectively.  The Wilson line for $n$-collinear
fields has the form of
$$W_{n}(x) = \bigg[\sum_{\mathrm{perms}}\exp \bigg(\frac{-g}
{\bar\mathcal{{P}}}\bar{n}\cdot A_{n}(x) \bigg)\bigg],$$ which is
required to ensure collinear gauge invariance. The Lagrangian of
collinear sectors, which is invariant under the usoft and
collinear gauge transformation, at leading order\footnote{We'll
restrict our discussion only at this order through the paper.}
(LO) in $\lambda$ is \cite{bsr},
\begin{eqnarray}
\nonumber \mathcal{L}_{c}&=& \mathcal{L}_{cg}+\mathcal{L}_{cq}, \nonumber\\
\mathcal{L}_{cg}&=&
\frac{1}{2g^2}\mathrm{Tr}\{[i\mathcal{D}^{\mu}+gA_{n}^{\mu},i\mathcal{D}^{\nu}
+gA_{n}^{\nu}]\}^2 \nonumber \\
&&+2\mathrm{Tr}\{\bar{c}_{n}[i\mathcal{D}_{\mu},[i\mathcal{D}^{\mu}+
A^{\mu}_{n},c_{n}]]\}+\frac{1}{\alpha}
\mathrm{Tr}\{[i\mathcal{D}_{\mu},A_{n}^{\mu}]\}^2,\\
\mathcal{L}_{cq}&=&\bar{\xi}_{n}[in\cdot
D+i\not\!\!{D}_{\perp}^c\frac{1}{i\bar{n}\cdot
D^c}i\not\!\!{D}_{\perp}^c] \frac{\not\!{\bar{n}}}{2}\xi_{n}.
\end{eqnarray}
Here the third line are the gauge fixing terms with parameter
$\alpha$ and $c_{n}$ denotes collinear ghost field, and
\begin{eqnarray}
\nonumber i\mathcal{D}^{\mu}&=&\bar\mathcal{P}\frac{n^\mu}{2}+
\mathcal{P}_{\bot}^\mu+(in\cdot\partial+gn\cdot A_{us})
\frac{\bar{n}^\mu}{2}, \nonumber\\
in\cdot D &=& in\cdot D_{us}+ g n\cdot A_n, \qquad\
iD_{us}=i\partial+gA_{us}, \nonumber\\
i\bar{n}\cdot D^c &=& \bar\mathcal{{P}}+g\bar{n}\cdot A_n, \qquad\
iD_{\perp}^c=\mathcal{P}_{\bot}+gA_n^\perp.
\end{eqnarray}The Lagrangian of soft sectors in SCET is identical to that of
QCD.

As for SCET$_\mathrm{II}$, it was emphasized that it can also be
viewed as the EFT of SCET$_\mathrm{I}$, and is the final theory
\cite{rpi}. This suggests a short path to go into
SCET$_\mathrm{II}$ from SCET$_\mathrm{I}$, if the following
matching and running steps are taken \cite{rpi}:\\
(1) Matching QCD onto SCET$_{\mathrm{I}}$ at a scale $\mu^{2}\sim
Q^{2}$ with $p_{c}^{2}\sim Q^{2}\lambda^{2}$;\\
(2) Decoupling the usoft-collinear interactions with the field
redefinitions, $\xi_{n}=Y_{n}^{\dagger}\xi_{n}^{(0)}$ and
$A_{n}=Y_{n}^{\dag}A_{n}^{(0)}Y_{n}$. Here
$Y_{n}(x)=\mathrm{P}\exp(ig\int dsn\cdot A_{us}(ns+x))$ is the
usoft Wilson line of usoft gluons in $n$ direction from $s=0$ to
$s=\infty$ for final state particles, and $\mathrm{P}$ means
path-ordered product, while for initial state particles, $Y_{n}$
is from $s=-\infty$ to $s=0$ and the daggers are reversed. This
step leads to $\mathcal{L}_{c}(\xi_{n},A_n,n\cdot A_{us})
=\mathcal{L}_{c}(\xi_{n}^{(0)},A_n^{(0)},0)$;\\
(3) Matching SCET$_{\mathrm{I}}$ onto SCET$_{\mathrm{II}}$ at a
scale $\mu^{2}\sim Q^{2}\lambda^{2}$ with $p_{c}^{2}\sim
Q^{2}\eta^{2}$. Thus, the soft and collinear modes are decoupled
in the Lagrangian of SCET$_{\mathrm{II}}$.

Next, we extend SCET to include the possibility of collinear
fields moving in different light-cone directions
$n_1,n_2,n_3,...$. These directions defined by $n_{i}$ and $n_{j}$
satisfy $n_{i}\cdot n_{j}\gg\lambda^{2} $ for $i\neq j$. For
simplicity we will only consider the case of head-on jets
corresponding to collinear particles moving in the $n$ and
$\bar{n}$ directions. Since the effective theory only takes
account the interactions of the modes in the local way, the
Lagrangian of the effective theory contains no direct coupling of
collinear particles moving in the two separate directions, however
the usoft gluons can mediate between them in SCET$_\mathrm{I}$.
Hence the Lagrangian in this case can be written by
$$\mathcal{L}^{c}_{\{n,\bar{n}\}}=\mathcal{L}^{c}_{n}+\mathcal{L}^{c}_{\overline{n}},$$
and the soft parts are unchanged, so the decoupling
transformations are also valid here.

To illustrate the application of SCET and warm up, we consider the
Sudakov effect of quark electromagnetic form factor in QCD
\cite{onfsf}, i.e., the double logarithmic asymptotic of
conservative current $j^{\mu}=\bar{\psi}\gamma^{\mu}\psi$ in the
following kinematics:\\
(a) nearly on-shell case
\begin{eqnarray}
&& Q^{2}=-(p_{1}-p_{2})^{2}\gg -p_{1}^{2}=-p_{2}^{2} \sim
\Lambda_{QCD}^{2},\nonumber \\
&& p_{1}\sim Q(\eta^{2},1,\eta), \qquad\ p_{2}\sim
Q(1,\eta^{2},\eta), \qquad\ \eta\sim \Lambda_{QCD}/Q;\nonumber
\end{eqnarray}
(b) off-shell case
\begin{eqnarray}
&& Q^{2}=-(p_{1}-p_{2})^{2}\gg -p_{1}^{2}=-p_{2}^{2}\sim
Q\Lambda_{QCD},\nonumber \\
&& p_{1}\sim Q(\lambda^{2},1,\lambda), \qquad\ p_{2}\sim
Q(1,\lambda^{2},\lambda), \qquad\ \lambda\sim
\sqrt{\Lambda_{QCD}/Q}.\nonumber
\end{eqnarray} Here $p_{1},p_{2}$ are the
momenta of the initial and final quarks. In the above two cases we
have omitted quark mass effects.

Following the treatment of heavy-to-collinear current discussed in
\cite{rpi,hl} for case (a), we first match the full current onto
the corresponding operator in SCET$_{\mathrm{I}}$. At LO in
$\lambda$, it gives \cite{ee}
\begin{equation}
\label{m} j^{\mu}=[\bar{\xi}_{\bar{n}}W_{\bar{n}}]\gamma^{\mu}
C_{q}(\mathcal{P}^\dagger,\bar\mathcal{{P}},\mu^{2})[W_{n}^{\dagger}\xi_{n}].
\end{equation}
By the requirement of collinear and usoft gauge invariance, the LO
effective operator is determined uniquely, and the
re-parameterization invariance(RPI) \cite{rpi} implies that the
Wilson coefficient satisfies
$C_{q}(\mathcal{P}^\dagger,\bar\mathcal{{P}},\mu^{2})
=C_{q}(\mathcal{P}^\dagger\cdot\bar\mathcal{{P}},\mu^{2}).$

\begin{figure}[t!]
\includegraphics{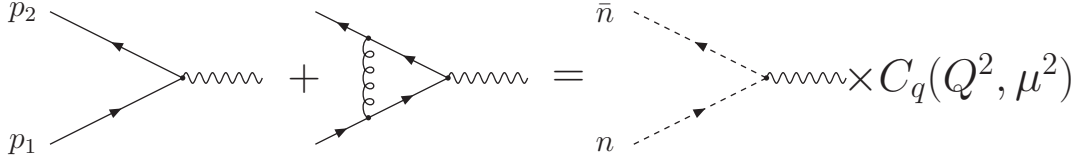}
\caption{\label{qqr} Graphical representation for quark current
matching.}
\end{figure}

Obviously, the tree level matching condition for $j^{\mu}$ leads
to $C_{q}(Q^{2},\mu^{2})=1+\mathcal{O}(\alpha_{s})$. We certainly
can determine $\mathcal{O}(\alpha_{s})$ correction by adopting
dimensional regularization\footnote{The $\overline{\mathrm{MS}}$
scheme, i.e., $d=4-2\epsilon$ and $\mu^2\rightarrow
\mu^2e^{\gamma_E}/4\pi$ is used through this paper, where
$\gamma_{E}$ is Euler's constant.} (DR) to regulate UV and IR
divergences to compute on-shell matrix elements on both sides of
Eq.(\ref{m}), for which is valid at the operator level and the
matching calculation is independent of regularization method. With
this choice, the fact
$\mathrm{IR}_{\mathrm{\mathrm{QCD}}}=\mathrm{IR}_{\mathrm{SCET}}$
provides us a direct matching calculation to read off the Wilson
coefficients and anomalous dimensions of the operators in SCET. As
all the on-shell loop integrals in SCET are scaleless and vanish,
$\mathrm{IR}_{\mathrm{SCET}} =-\mathrm{UV}_{\mathrm{SCET}}$, and
then
$\mathrm{IR}_{\mathrm{\mathrm{QCD}}}=-\mathrm{UV}_{\mathrm{SCET}}$.
Furthermore, the self energy diagrams in full QCD with massless
quarks are also vanish, and all the wave function renormalization
constants and the residues of the related propagators are equal to
unity, and we also note that the conserved current in full QCD
needs not to be renormalized. Thus, what we need to calculate is
an one particle irreducible diagram, Fig.\ref{qqr}, in the full
theory, which is given by
\begin{eqnarray}
\nonumber \langle p_2|j^{\mu}|p_1 \rangle
& = & \langle p_2|j^{\mu}|p_1 \rangle^{tree}+\langle p_2|j^{\mu}|p_1 \rangle^{one-loop} \nonumber \\
& = & \bar{u}(p_2)\gamma^{\mu}(1+V_{q})u(p_1),
\end{eqnarray}
\begin{eqnarray}
\label{qq} V_{q} & = & \frac{\alpha_{s}C_{F}}{4\pi}
(\frac{\mu^{2}e^{\gamma_{E}}}{Q^{2}})^{\epsilon}
\frac{\Gamma(1-\epsilon)}{\Gamma(1-2\epsilon)}
(-\frac{2}{\epsilon^{2}}-\frac{3}{\epsilon}
-8-\frac{\pi^{2}}{3}) \nonumber \\
& = & \frac{\alpha_{s}C_{F}}{4\pi}
\bigg[-\frac{2}{\epsilon^{2}}-\frac{1}{\epsilon}
(2\log\frac{\mu^2}{Q^2}+3) \nonumber \\
&& - \log^{2}\frac{\mu^2}{Q^2}-3\log\frac{\mu^2}{Q^2}
-8+\frac{\pi^{2}}{6}\bigg].
\end{eqnarray}
Here $C_F=(N_c^2-1)/2N_c$ for $SU(N_c)$, and $N_c=3$ for QCD. The
$\epsilon$-poles in Eq.(\ref{qq}) are of IR character, whose
opposition are just the UV poles in SCET. Thus the matching
calculation at one-loop level gives
\begin{eqnarray}
&& C_{q}(Q^{2},Q^{2})
 = 1+\frac{\alpha_{s}C_{F}}{4\pi}(-8+\frac{\pi^{2}}{6}),\\
\label{uvq} && Z_{V} \equiv \sum_{n}\frac{Z_{V}^{(n)}}{\epsilon^{n}} \nonumber \\
&& \hspace{0.7cm} =  1+\frac{\alpha_{s}C_{F}}{4\pi} \left
[\frac{2}{\epsilon^2}+\frac{1}{\epsilon}(2\log\frac{\mu^2}{Q^2}+3)
\right ].
\end{eqnarray}
Here $Z_{V}$ is defined as $\overline{\mathrm{MS}}$
renormalization constant of the effective operator, and $\mu$ has
been set to $Q$ to minimize the logarithms in the Wilson
coefficient. It was pointed out \cite{bsr} that the anomalous
dimension of the effective operator is independent of its spin
structure, for which can be factorized out from loop integrals.
This means the evolution equation in SCET is universal, and only
the Wilson coefficient is process-dependent.

From Eq.(\ref{uvq}), we obtain the RGE of $C_{q}(Q^{2},\mu^2)$,
\begin{eqnarray}
&& \label{rgq} \frac{d\log C_{q}(Q^{2},\mu^2)}{d\mathrm{log}(\mu)}
= \gamma_{1}(\mu) = -g \frac{\partial Z_{V}^{(1)}}{\partial g}, \\
&& \gamma_{1}(\mu) \equiv \mathbf{A}_{q}(\alpha_{s})\log
\frac{Q^{2}}{\mu^{2}}+ \mathbf{B}_{q}(\alpha_{s}) \nonumber \\
&& \hspace{1.1cm} = -\frac{\alpha_{s}C_{F}}{4\pi}
(4\log\frac{\mu^2}{Q^2}+6) .
\end{eqnarray}
Here $\mathbf{A}_{q}^{(1)}=C_F $ and
$\mathbf{B}_{q}^{(1)}=-\frac{3}{2}C_F$\footnote{The notion
$\mathbf{A}\equiv
\sum_{n}({\alpha_{s}}/{\pi})^{n}\mathbf{A}^{(n)},$ we adopt
$\{\mathbf{A},\mathbf{B}\}$ to distinguish the well known
coefficients $\{A, B\}$ in pQCD.}. With Eq.(\ref{rgq}), we can
resum the terms such as double logarithms from the scale $\sim
Q^{2}$ down to the scale $\sim Q^{2}\lambda^{2}$, we abbreviate
this matching step as a chain $QCD|_{Q^{2}}\longrightarrow
SCET_I|_{Q^{2}\lambda^{2}}$.

Next, We decouple the usoft and collinear modes by the field
redefinitions, which results in
\begin{equation}
\langle p_2|[\bar{\xi}_{\bar{n}}W_{\bar{n}}]\gamma^{\mu}
[W_{n}^{\dagger}\xi_{n}]|p_1\rangle\longrightarrow\langle
p_2|[\bar{\xi}_{\bar{n}}^{(0)}W_{\bar{n}}^{(0)}]|\Omega\rangle\gamma^{\mu}
\langle\Omega|\mathrm{T}[Y_{\bar{n}}Y_{n}]|\Omega\rangle
\langle\Omega|[W_{n}^{(0)\dagger}\xi_{n}^{(0)}]|p_1\rangle,
\end{equation}
where $\mathrm{T}$ means time ordering operator.

For the final step we integrate out all the off-shell modes of
order $\sqrt{Q\Lambda_{QCD}}$ and go into SCET$_{\mathrm{II}}$. We
can rename the usoft fields as soft fields for the usoft degrees
of freedom scaling as soft ones, and then lower the the
off-shellness of the collinear fields that would be matched onto
SCET$_{\mathrm{II}}$. Since the leading collinear Lagrangians in
SCET$_{\mathrm{I}}$ and SCET$_{\mathrm{II}}$ are the same and
(u)soft and collinear fields are decoupled at LO in
$(\lambda)\eta$, all possible time-ordered products involve
collinear fields agree exactly and we can simply replace
$\bar{\xi}_{\bar{n}}^{(0)}W_{\bar{n}}^{(0)}\rightarrow
\bar{\xi}_{\bar{n}}^{\mathrm{II}}W_{\bar{n}}^{\mathrm{II}}$ and
$W_{n}^{(0)\dagger}\xi_{n}^{(0)}\rightarrow
W_{n}^{\mathrm{II}\dagger}\xi_{n}^{\mathrm{II}}$, where the
superscript II denotes SCET$_{\mathrm{II}}$ will be dropped from
now on.

\begin{figure}[t!]
\begin{minipage}{0.45\textwidth}
\includegraphics{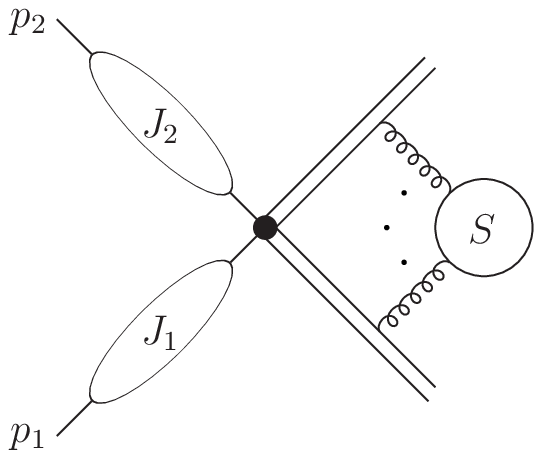}\\
(a)
\end{minipage}
\begin{minipage}{0.45\textwidth}
\includegraphics{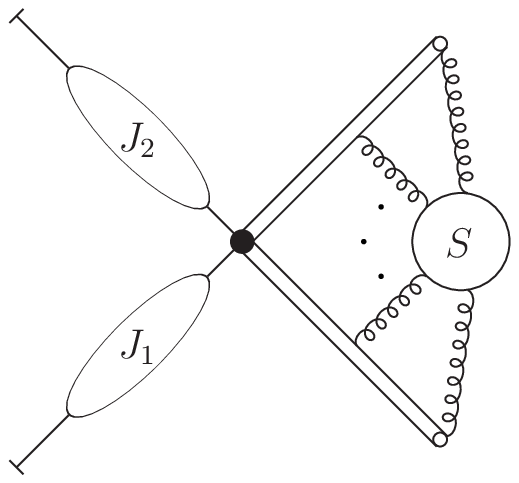}\\
(b)
\end{minipage}
\caption{\label{offshell}Factorization of on-shell form factor (a)
and off-shell form factor (b) in SCET. In (b), the soft Wilson
lines are terminated and the external quarks are amputated,
therefore, (b) can be taken as a sub-diagram of (a). Both diagrams
are depicted under the gauge $ \bar{n} \cdot A_{n}=n \cdot
A_{\bar{n}} = 0 $.}
\end{figure}
Because $p_1,p_2$ of (a) are described by the collinear modes in
SCET$_{\mathrm{II}}$, the general matching structure of
SCET$_{\mathrm{II}}$ diagram is shown in Fig.2.(a). The Wilson
coefficient at this step is unity and anomalous dimension is the
same as the first step, except that it runs from the scale $\sim
Q^{2}\lambda^{2}$ to the scale $\sim Q^{2}\eta^{2}$. We abbreviate
this step as $SCET_{I}|_{Q^{2}\lambda^{2}}\Rightarrow
SCET_{II}|_{Q^{2}\eta^{2}}$. Collecting all the results above, we
obtain the known Sudakov form factor
$S_{q}^{(a)}(\Lambda_{\mathrm{QCD}},Q)$, leaving other
coefficients omitted,
\begin{equation}
S_{q}^{(a)}(\Lambda_{\mathrm{QCD}},Q)=\exp\bigg(-\int_{\Lambda_{\mathrm{QCD}}}^{Q}
\gamma_{1}(\mu)d\log\mu\bigg).
\end{equation}

For case (b), it can be taken as a sub-diagram of the on-shell
case, from kinematical considerations, of which the external legs
are amputated. Thus, step (1) is unchanged, and in step (2),
$\langle0|\mathrm{T}[Y_{\bar{n}}Y_{n}]|0 \rangle$ changes into
\begin{equation}
\label{wl} -\int_{0}^{\infty}\int_{0}^{\infty}dsdt
e^{iQ\lambda^{2}(s+t)}\langle
\Omega|\mathrm{T}[Y_{\bar{n}}(0,\bar{n}s)Y_{n}^{\dagger}(0,-nt)]|\Omega\rangle,
\end{equation}
where $1/(Q\lambda^{2})$ is the effective contour length
\cite{onfsf} and
\begin{eqnarray}
Y_{\bar{n}}(0,\bar{n}s) &\equiv& \mathrm{P}
\exp\bigg(ig\int_{0}^{s}d\beta\bar{n}\cdot A_{us}(\bar{n}\beta)\bigg), \\
Y_{n}(0,-nt) &\equiv& \mathrm{P}\exp\bigg(-ig\int_{0}^{t}d\beta
n\cdot A_{us}(-n\beta)\bigg).
\end{eqnarray}
Because of the jets $J_1,J_2$ with fluctuations
$-p_{1}^{2}=-p_{2}^{2}=Q\Lambda_{QCD}\gg Q^2\eta^{2}$ in
Fig.2.(b), they must be integrated out in SCET$_{\mathrm{II}}$,
and only (\ref{wl}) is left over after step (3) associated with
renaming the usoft modes in SCET$_{\mathrm{I}}$ as the soft modes
in SCET$_{\mathrm{II}}$, of which the running behavior is the same
as $F_{IR}$ of \cite{onfsf}. Finally, the Sudakov factor in the
off-shell case is
\begin{eqnarray}
S_{q}^{(b)}(Q\eta,Q)&=&\exp\bigg(-\int_{Q\lambda}^{Q}\gamma_{1}(\mu)d\log\mu-
\int_{Q\eta}^{Q\lambda}\gamma_{2}(\mu)d\log\mu\bigg),\\
\gamma_{2}(\mu)&=&\frac{\alpha_{s}}{\pi}C_{F}\log
\frac{\mu^{2}}{Q^{2}\eta^{2}}+\mathcal{O}(\alpha_{s}^2),
\end{eqnarray}
where $\gamma_{2}(\mu)$ is the anomalous dimension of (\ref{wl}).
We conclude this section with a chain for the off-shell case,
$QCD|_{Q^2}\longrightarrow
SCET_{I}|_{Q^{2}\lambda^{2}}\longrightarrow
SCET_{II}|_{Q^{2}\eta^{2}}$. Now we are ready to turn into the
$Q_{T}$ resummation in the following.

\section{Method of $Q_T$ resummation in SCET}

Since SCET is powerful to disentangle the soft and collinear
interaction, and IR power counting \cite{ir} tells that the
singular terms of $Q_T$ distribution for DY-like processes in the
limit of $Q_T\rightarrow 0$ originate from soft and collinear
modes, which are emitted by partons from hadrons $p_1,p_2$, it is
not unexpected that SCET can be applied to treat them and to
derive the resummed part of full transverse momentum distribution
for these semi-inclusive processes, while the remaining regular
terms $Y$ \cite{abc,kt,pt} and the prescription of incorporating
non-perturbative region ($Q_T \sim \Lambda_{QCD}$) are neglected
in this paper.

For the sake of simplicity, the process of Higgs-boson production
is taken as a demonstration, but the method we used is not
confined to this example. The dominant process for Higgs-boson
production at the Large Hadron Collider  (LHC) in the Standard
Model are gluon fusion through a heavy quark loop, mainly the top
quark, $p_1(P_{1})+p_2(P_{2})\rightarrow gg\rightarrow \phi(Q)+X$
with $P_{1}=(0,2p,0)$, $P_{2}=(2p,0,0)$ and
$S=P_{1}^{-}P_{2}^{+}$. It is convenient to start from the
effective Lagrangian for one Higgs-boson and gluons coupling
\cite{h},
\begin{equation}
\mathcal{L}_{\phi gg} =\tau(\alpha_s)\phi
G_{\mu\nu}^{a}G_{a}^{\mu\nu},
\end{equation}
where
$\tau(\alpha_{s})=\frac{\alpha_{s}(Q)}{12\pi}(\sqrt{2}G_F)^{1/2}+\mathcal{O}(\alpha_{s}^{2})$
and $Q = m_{\phi}$. Therefore, the operator for Higgs-boson
production is $\mathcal{H}=G_{\mu\nu}^{a}G_{a}^{\mu\nu}$. Here the
coupling $\alpha_{s}$ suffers the QCD correction, which is unlike
the case of electro-charge coupling. Furthermore, because the
renormalization constant of
$\alpha_{s}G_{\mu\nu}^{a}G_{a}^{\mu\nu}$ is unity up to
$\mathcal{O}(\alpha_{s})$, the renormalization constant of
$\mathcal{H}$ is just $Z_{g}^{-2}$, where $Z_{g}$ is the
renormalization constant of gauge coupling-$g$.

If we set $\lambda^{2}\sim{Q_T}/{Q}$ with $Q\gg Q_T\gg
\Lambda_{QCD}$, the situation is much like that of quark form
factor-(a) discussed in last section, and the matching and running
procedure can be followed. The operator $\mathcal{H}$ can match at
LO in $\lambda$ onto
\begin{eqnarray}
\mathcal{H}&=&\frac{1}{2}\mathcal{B}_{\bar{n}a}^{\mu}
C_{g}(\mathcal{P}^\dagger\cdot\bar\mathcal{P},\mu^{2})
\mathcal{B}_{n\mu}^a,\nonumber \\
&=&\frac{1}{2} \mathcal{B}_{\bar{n}a}^{\mu} C_{g}(Q^2,\mu^{2})
\mathcal{B}_{n\mu}^a,
\end{eqnarray}where
\begin{eqnarray}
\mathcal{B}_{n}^{\mu}&=& \bar{n}_{\nu}\mathcal{G}_{n}^{\nu\mu},\nonumber \\
\mathcal{G}_{n}^{\mu\nu}&=&W_{n}^{\dagger}[i\mathcal{D}_{n}^{\mu}+gA_{n}^{\mu},
i\mathcal{D}_{n}^{\nu}+gA_{n}^{\nu}]W_{n},
\end{eqnarray}
and $n\leftrightarrow \bar{n}$ for
$\mathcal{G}_{\bar{n}}^{\mu\nu}$.

\begin{figure}[t!]
\includegraphics{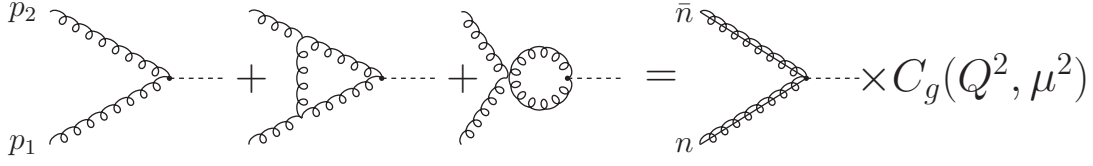}
\caption{\label{ggh} Graphical representation for gluon current
matching.}
\end{figure}
The one loop calculation, Fig.\ref{ggh}, is similar to quark
current, except for dividing the final result by $Z_g^{-2}$.
Finally,
\begin{eqnarray}
\nonumber \langle g_{1}|\mathcal{H}|g_{2} \rangle &=& \langle
g_{1}|\mathcal{H}|g_{2} \rangle^{tree}+\langle
g_{1}|\mathcal{H}|g_{2} \rangle^{one-loop}+c.t. \nonumber \\
&=& \langle g_{1}|\mathcal{H}|g_{2} \rangle^{tree}(1+V_{g}),
\end{eqnarray}
\begin{eqnarray} V_{g} &=&
\frac{\alpha_{s}}{4\pi}\left[
\frac{\Gamma(1-\epsilon)}{\Gamma(1-2\epsilon)}
\left(\frac{\mu^{2}e^{\gamma_{E}}}{Q^{2}}\right)^{\epsilon}
\left(-\frac{2C_{A}}{\epsilon^{2}}-\frac{2\beta_{0}}
{\epsilon}+\mathcal{A}_{g}^{H}\right)\right] \nonumber \\
&=& \frac{\alpha_{s}}{4\pi}
\bigg[-\frac{2C_{A}}{\epsilon^{2}}-\frac{1}{\epsilon}
(2C_{A}\log\frac{\mu^2}{Q^2}+2\beta_{0}) \nonumber \\
&& -C_{A}\log^{2}\frac{\mu^2}{Q^2}-2\beta_{0}\log
\frac{\mu^2}{Q^2}
+\mathcal{A}_{g}^{H}+\frac{C_{A}\pi^{2}}{2}\bigg]
\end{eqnarray}is obtained\footnote{Here we have absorbed the scale dependence
of $\alpha_s$ into that of $Z_g$.}, from which we read
\begin{eqnarray}
&& C_{g}(Q^2,Q^2) = 1+\frac{\alpha_{s}}{4\pi}
(\mathcal{A}_{g}^{H}+\frac{C_A\pi^{2}}{2}),\\
&& Z_{\mathcal{H}} \equiv
\sum_{n}\frac{Z_{\mathcal{H}}^{(n)}}{\epsilon^{n}} =
1+\frac{\alpha_{s}}{4\pi}
\bigg[\frac{2C_{A}}{\epsilon^{2}}+\frac{1}{\epsilon}(
2C_{A}\log\frac{\mu^{2}}{Q^{2}}+2\beta_{0})\bigg].
\end{eqnarray}
Here $C_A=N_c$, $\beta_{0}=\frac{11}{6}C_A-\frac{2}{3}n_fT_R$,
$\mathcal{A}_{g}^{H}=11+2\pi^{2}$, $T_R=\frac{1}{2}$ and $n_f=5$
is the number of active quark flavors. Then the RGE of
$C_g(Q^2,\mu^2)$ is
\begin{eqnarray}
\label{rgg}&& \frac{d\log C_{g}(Q^2,\mu^{2})}{d\log(\mu)}=
\gamma_{1}(\mu) = -g \frac{\partial Z_{\mathcal{H}}^{(1)}}{\partial g},\\
&& \gamma_{1}(\mu) \equiv \mathbf{A}_{g}(\alpha_{s})\log
\frac{Q^{2}}{\mu^{2}}+\mathbf{B}_{g}(\alpha_{s})\nonumber \\
&& \hspace{1.1cm} = -(\frac{\alpha_{s}}{\pi})
(C_{A}\log\frac{\mu^2}{Q^2}+\beta_{0}),
\end{eqnarray}
with $\mathbf{A}_{g}^{(1)}=C_A $ and
$\mathbf{B}_{g}^{(1)}=-\beta_{0}$. Thus the evolution from the
scale $\sim Q^{2}$ to the scale $\sim Q^{2}\lambda^{2}$ gives
\begin{equation}
C_{g}(Q^2,Q^{2}\lambda^{2})=
C_{g}(Q^2,Q^2)\exp\bigg(-\int_{Q^{2}\lambda^{2}}^{Q^{2}}
\frac{d\mu^{2}}{2\mu^{2}}\gamma_{1}(\mu)\bigg).
\end{equation}

As shown above, the extraction of $\mathbf{A},\mathbf{B}$ in SCET
is different from that of $A,B$ in pQCD, i.e., there is no need to
calculate real correction which is more difficult to handle. Using
the virtual part of higher order calculation, such as the two loop
on-shell quark and gluon form factor \cite{ab2}, we can find the
$\mathcal{O}(\alpha_{s}^{2})$ universal anomalous dimension. For
example,
\begin{eqnarray}
\mathbf{A}_{a}^{(2)}&=& \frac{1}{2}C_a K,\qquad\ K=
C_A(\frac{67}{18}-\frac{\pi^{2}}{6})-\frac{10}{9}T_{R}n_{f},
\end{eqnarray}where $C_q=C_{F}$, $C_g=C_A$.

After performing field redefinitions, Eq.(\ref{rgg}) can be
directly used to running $\mathcal{H}$ from the scale $\sim Q^{2}$
to the scale $\sim Q^{2}\eta^{2}$ without loss of degrees of
freedom. So the relevant operator for Higgs-boson production at
the scale $\sim Q^{2}\eta^{2}(Q_T^{2})$ is
\begin{eqnarray} \mathcal{H}&=&
C_{g}(Q^2,Q^2\eta^{2})\mathrm{Tr}
\{\mathrm{T}[Y_{n}^{\dagger}Y_{\bar{n}}\mathcal{B}_{\bar{n}}^{\mu}
Y_{\bar{n}}^{\dagger}Y_{n}\mathcal{B}_{n\mu}]\} \nonumber \\
&=& C_{g}(Q^2,Q^2\eta^{2}) \frac{1}{2}
\mathrm{T}[\mathcal{Y}_{\bar{n}}^{ab}\mathcal{Y}_n^{ac}
\mathcal{B}_{\bar{n}}^{b\mu}\mathcal{B}_{n\mu}^c]\nonumber \\
&\equiv& C_{g}(Q^2,Q^2\eta^{2})\hat{\mathcal{H}}.
\end{eqnarray}
Here $\mathcal{Y}_{n(\bar{n})}$ is the adjoint soft Wilson line
from $-\infty$ to $0$ in $n(\bar{n})$ direction for incoming
fields. Now, we have completed the procedures corresponding to
step (1), (2) and (3).

\begin{figure}[t!]
\includegraphics{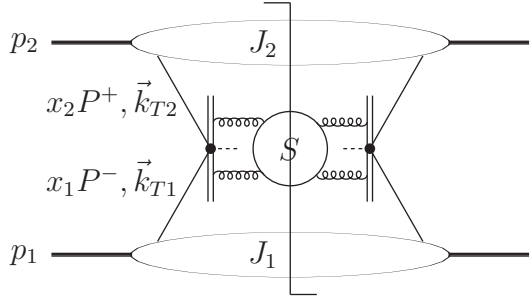}
\caption{\label{hpt}General structure of SCET cross section for
Higgs-boson production under the gauge $n\cdot
A_{\bar{n}}=\bar{n}\cdot A_n=0$}.
\end{figure}

To obtain the differential cross section, we relate it to the
composite operator $\mathcal{H}$ at the renormalization scale
$\mu^{2}\sim Q^{2}\eta^{2}$ in SCET$_{\mathrm{II}}$, where the
cross section can be written as
\begin{equation}
\label{q1}\frac{1}{\sigma_{gg}^{(0)}}\frac{{d\sigma}^{\mathrm{resum}}}
{dQ^{2}dydQ_{T}^{2}}=\mathbf{H}_{g}^{\phi}(Q)
e^{-\mathbf{S}_{g}(\mu,Q)}\sigma_{\mathrm{SCET}}(Q_T,Q,\mu),
\end{equation}
where $\mathbf{H}_{g}^{\phi}(Q)=|C_g(Q^2,Q^2)|^{2}$ is a function
of $\alpha_{s}(Q)$, and
\begin{eqnarray}
\sigma_{gg}^{(0)}&=&(\sqrt{2}G_{F})\frac{\alpha_{s}^{2}(Q)m_{H}^{2}}
{576 S}\delta(Q^{2}-m_{H}^{2}),\\
\mathbf{S}_{g}(\mu,Q)&=& \int_{\mu^{2}}^{Q^{2}}
\frac{d\mu^{2}}{\mu^{2}}
\bigg[\mathbf{A}_{g}(\alpha_{s})\mathrm{log}
\frac{Q^{2}}{\mu^{2}}+\mathbf{B}_{g}(\alpha_{s})\bigg],
\end{eqnarray}
and \begin{eqnarray}\sigma_{\mathrm{SCET}}(Q_T,Q,\mu)
=\frac{1}{\sigma_{gg}^{(0)}}\frac{d\sigma^{\mathrm{SCET}}(\mu)}{dQ^{2}dydQ_{T}^{2}}
\end{eqnarray}
represents the normalized differential cross section calculated in
SCET$_{\mathrm{II}}$ with the composite operator
$\hat{\mathcal{H}}$. The general structure of relevant diagram is
shown in Fig.~\ref{hpt}, where the soft and collinear modes are
decoupled and the spin and color are summed over in the matrix
element of hadron, from which the SCET cross section can be
written in the form of multiple convolution,
\begin{eqnarray}
\sigma_{\mathrm{SCET}}(Q_T,Q,\mu)&=& \int
d^{2}\vec{k}_{T1}d^{2}\vec{k}_{T2}d^{2}\vec{k}_{TS}
\delta^{2}(\vec{k}_{T1}+\vec{k}_{T2}+\vec{k}_{TS}-\vec{Q}_T) \nonumber\\
&& \times
J_{p_1}(x_{1},k_{T1},\mu)J_{p_2}(x_{2},k_{T2},\mu)S(k_{TS},\mu),
\end{eqnarray}
where $x_{1}={Qe^{y}}/ \sqrt{S}$, $x_{2}={Qe^{-y}}/\sqrt{S}$ for
$Q_{T}^{2} \ll Q^{2}$, and
\begin{eqnarray}
J_{p_1}(x_{1},k_{T1},\mu)&=& \frac{2}{x_{1}P^{-}}
\frac{1}{(2\pi)^{3}}\int dy^{+}d^{2}\vec{y}_{\perp}
e^{-i(x_{1}P^{-}y^{+}-\vec{k}_{T1}\cdot\vec{y}_{\perp})}\nonumber\\
&&\times\langle p_1
|\mathrm{Tr}[\mathcal{B}_{n}^{\alpha}(y^{+},0,\vec{y}_{\perp})
\mathcal{B}_{n\alpha}(0)]| p_1 \rangle,\\
J_{p_2}(x_{2},k_{T2},\mu)&=& \frac{2}{x_{2}P^{+}}
\frac{1}{(2\pi)^{3}}\int dy^{-}d^{2}\vec{y}_{\perp}
e^{-i(x_{2}P^{+}y^{-}-\vec{k}_{T2}\cdot\vec{y}_{\perp})}
\nonumber\\
&&\times\langle
p_2|\mathrm{Tr}[\mathcal{B}_{\bar{n}}^{\alpha}(0,y^{-},\vec{y}_{\perp})
\mathcal{B}_{\bar{n}\alpha}(0)]| p_2 \rangle,\\
S(k_{TS},\mu)&=& \frac{1}{(2\pi)^{2}}\int
d^{2}\vec{y}_{\perp}e^{i\vec{k}_{TS}\cdot\vec{y}_{\perp}}\langle
\Omega|\bar{\mathrm{T}}[\mathcal{Y}_n^{\dagger ec}
\mathcal{Y}_{\bar{n}}^{\dagger eb}](0,0,\vec{y}_{\perp}) \nonumber\\
&&\times\mathrm{T}[\mathcal{Y}_{\bar{n}}^{ab}\mathcal{Y}_n^{ac}](0)|\Omega\rangle,
\end{eqnarray}with $\bar{\mathrm{T}}$ denoting the anti-time ordering operator.
Obviously, in Eq.(38) the matrix element has been factorized, and
the delta function is imposed by momentum conservation.

Next, to factorize the phase space, the trick of Fourier
transforming to impact parameter space is significant \cite{abc},
\begin{eqnarray}
\int d^{2}\vec{Q}_T e^{i \vec{b}\cdot
\vec{Q}_T}\delta^{2}(\sum_{i}\vec{k}_{Ti}-\vec{Q}_T)=\prod_{i}e^{i
\vec{b}\cdot\vec{k}_{Ti}}.
\end{eqnarray}
Then, for each transverse momentum $\vec{k}_{Ti}$, one obtains
\begin{eqnarray}
\int d^{2}\vec{k}_{Ti} e^{i \vec{b}\cdot
\vec{k}_{Ti}}f(\vec{k}_{Ti})=\tilde{f}(b).
\end{eqnarray}
This produces the simple product
\begin{eqnarray}
\tilde{\sigma}_{\mathrm{SCET}}(b,Q,\mu)=
\tilde{J}_{p_1}(x_{1},b,\mu)\tilde{J}_{p_2}(x_{2},b,\mu)\tilde{S}({b},\mu).
\end{eqnarray}

Because of KLN theorem, the contributions from the soft modes are
free of IR divergences. So only the collinear divergences are
survived, therefore after matching the SCET cross section onto a
product of two parton distribution functions (PDFs) given by
\cite{hsf}, which are equivalent to the conventional PDFs
$f_{a/p_i}(x_{i},\mu)$ at LO in $\lambda$, the remaining IR
divergences can be absorbed into these nonperturbative inputs, of
which the evolutions are controlled by the DGLAP equations. This
leads to \cite{abc}
\begin{eqnarray}
\tilde{J}_{p_{i}}(x_i,b,\mu)&=& \sum_{a}(f_{a/p_{i}}\otimes
\mathbf{c}_{ga})(x_{i},b,\mu) \nonumber \\
&=& \sum_{a} \int_{x_{i}}^{1}\frac{d\xi}{\xi}
f_{a/p_{i}}(\xi,\mu) \mathbf{c}_{ga}(\frac{x_{i}}{\xi},b,\mu), \\
\mathbf{c}_{ga}&\equiv& \sum_{n=0}(\frac{\alpha_{s}}{4\pi})^{n}
\mathbf{c}_{ga}^{(n)}. \nonumber
\end{eqnarray}
If we define
\begin{eqnarray}
\mathbf{C}_{ga}(z,\frac{b_{0}}{b},\mu)
=\mathbf{c}_{ga}(z,b,\mu)[\tilde{S}({b},\mu)]^{\frac{1}{2}},
\qquad\ b_{0}=2e^{-\gamma_{E}},
\end{eqnarray}
then
\begin{eqnarray}
\label{q2}\tilde{\sigma}_{\mathrm{SCET}}(b,Q,\mu) =
(f_{a/p_{1}}\otimes
\mathbf{C}_{ga})(x_{1},\frac{b_{0}}{b},\mu)(f_{b/p_{2}}\otimes
\mathbf{C}_{gb})(x_{2},\frac{b_{0}}{b},\mu).
\end{eqnarray}
Obviously, $\mathbf{C}_{ga}^{(0)}(z)=\delta_{ga}\delta(1-z)$, and
we have derived in Appendix that
\begin{eqnarray}
\mathbf{C}_{ga}^{(1)}(z)&=&-2P_{ga}^{\epsilon}(z)
-C_{A}\frac{\pi^{2}}{6}\delta_{ga}\delta(1-z),
\end{eqnarray}
where $\mu$ has been set to $b_{0}/b$ to eliminate large constant
factors in $\mathbf{C}_{ga}^{(1)}(z,\mu)$, and
$P_{ga}^{\epsilon}(z)$ represent the $\mathcal{O}(\epsilon)$ terms
of the DGLAP splitting kernels.

Combining Eq.(\ref{q1})-(\ref{q2}) and Fourier transforming back
to $Q_T$ space, we obtain the resumed formula of transverse
momentum distribution for Higgs-boson production in SCET,
\begin{eqnarray}
\frac{1}{\sigma_{gg}^{(0)}}\frac{{d\sigma}^{\mathrm{resum}}}
{dQ^{2}dydQ_{T}^{2}}&=& \mathbf{H}_{g}^{\phi}(Q) \int_{0}^{\infty}
\frac{db}{2\pi} bJ_0(bQ_T)\sum_{ab}
e^{-\mathbf{S}_{g}(\frac{b_{0}}{b},Q)}\nonumber\\
&&\times(f_{a/p_{1}}\otimes
\mathbf{C}_{ga})(x_{1},\frac{b_{0}}{b})(f_{b/p_{2}}\otimes
\mathbf{C}_{gb})(x_{2},\frac{b_{0}}{b}).
\end{eqnarray}

The similar reasoning leads to the general form for $Q_T$
resummation,
\begin{eqnarray}
\label{eqt}
\frac{1}{\sigma^{(0)}}\frac{{d\sigma}^{\mathrm{resum}}}
{dQ^{2}dydQ_{T}^{2}}&=& \mathbf{H}_{c}^{F}(Q) \int_{0}^{\infty}
\frac{db}{2\pi} bJ_0(bQ_T)\sum_{ab}
e^{-\mathbf{S}_{c}(\frac{b_{0}}{b},Q)}\nonumber\\
&&\times (f_{a/p_{1}}\otimes
\mathbf{C}_{ca})(x_{1},\frac{b_{0}}{b})(f_{b/p_{2}}\otimes
\mathbf{C}_{\bar{c}b})(x_{2},\frac{b_{0}}{b}),
\end{eqnarray}where $F$ and $c$ stand for the type
of process and of parton participating the elementary sub-process,
for example, $F=DY$ and $c=q$ for Drell-Yan process. The formula
(\ref{eqt}) is a little different from the known CSS formula
\cite{abc},
\begin{eqnarray}
\label{css}\frac{1}{\sigma^{(0)}}\frac{{d\sigma}^{\mathrm{resum}}}
{dQ^{2}dydQ_{T}^{2}}&=& \int_{0}^{\infty} \frac{db}{2\pi}
bJ_0(bQ_T)\sum_{ab}e^{-S_{c}(\frac{b_{0}}{b},Q)}\nonumber\\
&&\times(f_{a/p_{1}}\otimes
C_{ca})(x_{1},\frac{b_{0}}{b})(f_{b/p_{2}}\otimes
C_{\bar{c}b})(x_{2},\frac{b_{0}}{b}).
\end{eqnarray}

Transforming Eq.(\ref{eqt}) into the form as Eq.(\ref{css}) by the
identity
\begin{eqnarray}
\mathbf{H}_{c}^{F}(Q)=\exp\left[\int^{Q^2}_{b_{0}^{2}/b^{2}}
\frac{d\mu^{2}}{\mu^{2}}\beta(\alpha_{s})\frac{d \log
\mathbf{H}_{c}^{F}}{d \log
\alpha_{s}}\right]\mathbf{H}_{c}^{F}({b_{0}/b})
\end{eqnarray} with $\beta(\alpha_{s})$ denoting the QCD
$\beta$-function, one finds
\begin{eqnarray}
\frac{1}{\sigma^{(0)}}\frac{{d\sigma}^{\mathrm{resum}}}
{dQ^{2}dydQ_{T}^{2}}&=& \int_{0}^{\infty} \frac{db}{2\pi}
bJ_0(bQ_T)\sum_{ab}e^{-\bar{S}_{c}(\frac{b_{0}}{b},Q)}\nonumber\\
&&\times(f_{a/p_{1}}\otimes
\bar{C}_{ca})(x_{1},\frac{b_{0}}{b})(f_{b/p_{2}}\otimes
\bar{C}_{\bar{c}b})(x_{2},\frac{b_{0}}{b}).
\end{eqnarray}
The corresponding coefficients $\{\bar{A},\bar{B},\bar{C}\}$ are
related to $\{\mathbf{A},\mathbf{B},\mathbf{C}\}$ through equation
\begin{eqnarray}
\nonumber \bar{A}(\alpha_{s}) &=& \mathbf{A}(\alpha_{s}),\nonumber\\
\bar{B}(\alpha_{s}) &=&
\label{rela}\mathbf{B}(\alpha_{s})-\beta(\alpha_{s})\frac{d \log
\mathbf{H}_{c}^{F}}{d \log \alpha_{s}},\\
\bar{C}_{ab}(z) &=& \mathbf{C}_{ab}(z)
[\mathbf{H}_{a}^{F}(b_{0}/b)]^{\frac{1}{2}},\nonumber
\end{eqnarray}
Up to next-to-leading-logarithmic-order (NLLO),
$\{\mathbf{A}^{(2)},\mathbf{B}^{(1)},\mathbf{C}^{(1)}\}$ and
$\{A^{(2)},B^{(1)},C^{(1)}\}$ is compatible with each other, and
further calculation and confirmation are required at higher order.

It can be seen that SCET provides a natural framework of $Q_T$
resummation by conventional RGE in EFT. We have noted that the
matched effective operator is determined by collinear and soft
gauge invariance and is unique. In addition, the corresponding
anomalous dimension is independent of its spin structure and is
universal, while the process-dependent quantity resides in the
Wilson coefficient. Even more, the coefficients $\mathbf{C}_{ab}$
defined in SCET are process-independent too. So the matching and
running procedure in EFT naturally separated the process-dependent
and universal contributions to a process, i.e.,
$\{\mathbf{A},\mathbf{B},\mathbf{C}\}$ are universal and only
$\mathbf{H}_{c}^{F}$ is process-dependent. In pQCD, the formula
and relation like Eq.(\ref{eqt}) and Eq.(\ref{rela}) have been
proposed by the authors of \cite{upt}.

Compared with SCET, pQCD analysis invoke gauge invariance and a
new evolution equation \cite{abc} which comes from differentiating
the jet-function from the factorized cross section with respect to
the axial parameter in axial gauge to separate soft and collinear
contributions, which is crucial to resum the double logarithms,
since RGE in full theory only resums single logarithms between two
scales.

We conclude this section with a chain for the $Q_T$ resummation in
this section,
$$QCD|_{Q^{2}}\longrightarrow SCET_{I}|_{Q^{2}\lambda^{2}}\Rightarrow
SCET_{II}|_{Q^{2}\eta^{2}}\longleftarrow DGLAP|_{\mu_{0}^{2}},$$
where the last arrow indicates that the PDFs used at the scale
$Q^{2}\eta^{2}$ can be obtained from those at some fixed scale
$\mu_{0}^{2}$ by the evolutions of the DGLAP equations.

\section{discussion}

(I) Applying the formula (\ref{eqt}) to the production of lepton
pair via virtual photon, one can find,
\begin{eqnarray}
\frac{1}{\sigma_{qq}^{(0)}}\frac{{d\sigma}^{\mathrm{resum}}}
{dQ^{2}dydQ_{T}^{2}}&=&\mathbf{H}_{q}^{DY}(Q)\int_{0}^{\infty}
\frac{db}{2\pi} bJ_0(bQ_T)\sum_{ab}
e^{-\mathbf{S}_{q}(\frac{b_{0}}{b},Q)}\nonumber\\
&&\times(f_{a/p_{1}}\otimes
\mathbf{C}_{qa})(x_{1},\frac{b_{0}}{b})(f_{b/p_{2}}\otimes
\mathbf{C}_{qb})(x_{2},\frac{b_{0}}{b}),
\end{eqnarray}
where
\begin{eqnarray}
\nonumber && \sigma_{qq}^{(0)} =
e_q^2\frac{4\pi^2\alpha_{em}^2}{9SQ^2},\qquad\
\mathbf{H}_{q}^{DY}(Q)=1+\frac{\alpha_{s}C_{F}}{2\pi}
(-8+\frac{7}{6}\pi^{2}),\nonumber\\
&& \mathbf{S}_{q}(\frac{b_{0}}{b},Q) =
\int_{b_{0}^{2}/{b}^{2}}^{Q^{2}}
\frac{d\mu^{2}}{\mu^{2}}\bigg[\mathbf{A}_{q}(\alpha_{s})\mathrm{log}
\frac{Q^{2}}{\mu^{2}}+\mathbf{B}_{q}(\alpha_{s})\bigg],\\
&& \mathbf{C}_{qa}^{(0)} = \delta_{qa}\delta(1-z), \qquad\
\mathbf{C}_{qa}^{(1)}=-2P_{qa}^{\epsilon}(z)
-C_{F}\frac{\pi^{2}}{6}\delta_{qa} \delta(1-z).\nonumber
\end{eqnarray}

(II) The reformulation of joint resummation can also be made
straightforwardly in SCET. In fact, threshold resummation
\cite{dy} under $z=Q^2/S\rightarrow1$ for
${d\sigma^{\mathrm{resum}}}/{dQ^2}$ is performed in moment-$N$
space, and the relevant $\lambda^{2}=1/\bar{N}\sim 1-z$ with
$\bar{N}=e^{\gamma_{E}}N$. The conclusion of \cite{dy} can be
represented by a chain,
$$QCD|_{Q^{2}}\longrightarrow SCET_{I}|_{Q^{2}\lambda^{2}}\Rightarrow
SCET_{II}|_{Q^{2}\eta^{2}}\longleftarrow DGLAP|_{\mu_{0}^{2}}.$$
We observe that the two chains for $Q_T$ and threshold resummation
in SCET have identical structure. This suggests that we can do
threshold and $Q_T$ resummation for
${d\sigma^{\mathrm{resum}}}/{dQ^2dQ_T^2}$ simultaneously. The
relevant $\lambda^{2}\sim 1/\chi(\bar{N},\bar{b})$ with
$\bar{b}\equiv bQ /b_{0}$ is an interpolation of $\lambda^{2}\sim
1/\bar{N}$ and $\lambda^{2}\sim 1/\bar{b}$, let us say \cite{jr}
\begin{equation}
\chi(\bar{N},\bar{b})=\bar{b}+\frac{\bar{N}}{1+\rho\bar{b}/\bar{N}},
\qquad\ \rho=\frac{1}{4},
\end{equation}
which approaches to $\bar{N}$ for $\bar{b} \ll \bar{N}$ and to
$\bar{b}$ for $\bar{b}\gg \bar{N}$, respectively. The matching
steps for joint resummation then can be written as
$$QCD|_{Q^{2}}\longrightarrow SCET_{I}|_{Q^{2}/\chi}\Rightarrow
SCET_{II}|_{Q^{2}/\chi^{2}}\longleftarrow DGLAP|_{\mu_{0}^{2}},$$
which leads to similar result as Eq.(\ref{eqt}) corresponding to
that of \cite{jr}, and the Mellin transformed and jointly resummed
cross section follows,
\begin{eqnarray}
\frac{1}{\sigma^{(0)}}\frac{{d\sigma}^{\mathrm{resum}}(N)}
{dQ^{2}dQ_{T}^{2}}&=&\mathbf{H}_{c}^{F}(Q)\int_{0}^{\infty}
\frac{db}{2\pi} bJ_0(bQ_T)\sum_{ab}
e^{-\mathbf{S}_{c}(Q/\chi,Q)}\nonumber\\
&&\times\bigg(\mathbf{C}_{ca}(N)f_{a/p_{1}}(N,Q/\chi)\bigg)
\bigg(\mathbf{C}_{\bar{c}b}(N)f_{b/p_{2}}(N,Q/\chi)\bigg),
\end{eqnarray}
where $\phi(N)\equiv \int_0^1d\xi\xi^{N}\phi(\xi)$ for any
function $\phi(\xi)$ with $0\leq\xi\leq1$ is used.

\section{conclusion}

We have presented the method of $Q_T$ resummation in the framework
of SCET and given a simple correspondence between
$\{\mathbf{A},\mathbf{B},\mathbf{C}\}$ in SCET and the well known
coefficients $\{A,B,C\}$ in pQCD, with which the available
information is compatible. The equivalence of the two framework
can be confirmed by higher order computation. We have also shown
that the reformulation of joint resummation can be performed in
SCET directly. So any process, which is confined to the soft and
collinear regions by dynamics or kinematics, can be treated in
SCET following the steps outlined above.

\section*{Acknowledgements}
We would like to thank C.W.Bauer for his kind discussion at the
Conference of QCD and Hadron Physics held in Peking University,
June 2005. We also thank X.D.Ji for useful discussions. This work
is supported in part by the National Natural Science Foundation of
China, under grant Nos. 10421003 and 10575001, and the key Grant
Project of Chinese Ministry of Education, under the Grant No.
305001.\vspace{.5cm}\\
{\it Note added.} -After an expanded version of our manuscript was
completed, a new paper \cite{qt} by A. Idilbi, X.D. Ji, and F.
Yuan appeared, in which they also discussed the transverse
momentum distribution in $b$ space and derived a similar result to
ours.

\section*{Appendix.A}
In this appendix, the details of the calculation to extract
$\mathbf{C}^{(1)}$ in SCET are given explicitly.

Because $\mathbf{C}^{(1)}$ is related to the emission of a soft or
collinear gluon, and the phase space is already factorized in this
case, we will exploit a special form at $\mathcal{O}(\alpha_{s})$,
for which there is no need to cover the non-perturbative region
($Q_T\sim \Lambda_{QCD}$),
\begin{eqnarray}
\frac{1}{\sigma^{(0)}}\frac{{d\sigma}^{\mathrm{resum}}}
{dQ^{2}dydQ_{T}^{2}}&=& \mathbf{H}_{c}^{F}(Q)\frac{d}{dQ_{T}^{2}}
\bigg[e^{-\mathbf{S}_{c}(\mu,Q)}\hat{\sigma}_{\mathrm{SCET}}(Q_T,Q,\mu)\bigg],\\
\label{m1}
\hat{\sigma}_{\mathrm{SCET}}(Q_T,Q,\mu)&=&\int_{0}^{Q_{T}^{2}}dq_{T}^{2}
\frac{1}{\sigma^{(0)}}\frac{d\sigma^{\mathrm{SCET}}
(\mu)}{dQ^{2}dydq_{T}^{2}}\nonumber\\
&=&J_{p_1}(x_{1},Q_{T},\mu)J_{p_2}(x_{2},Q_{T},\mu)S(Q_{T},\mu).
\end{eqnarray}
The same reasoning as in section 3 leads to
\begin{eqnarray}
\label{m2}
J_{p_1}(x_{1},Q_{T},\mu)J_{p_2}(x_{2},Q_{T},\mu)S(Q_{T},\mu)&=&\sum_{a,b}
(f_{a/p_{1}}\otimes\hat{\mathbf{C}}_{ca})(x_{1},Q_{T},\mu) \nonumber \\
&&\times(f_{b/p_{2}}\otimes
\hat{\mathbf{C}}_{\bar{c}b})(x_{2},Q_{T},\mu).
\end{eqnarray}
Then we get the differential form for $Q_{T}$ resummation,
\begin{eqnarray}
\label{qt}\frac{1}{\sigma^{(0)}}\frac{{d\sigma}^{\mathrm{resum}}}
{dQ^{2}dydQ_{T}^{2}}&=&\mathbf{H}_{c}^{F}(Q)\frac{d}{dQ_{T}^{2}}
\sum_{ab}
e^{-\mathbf{S}_{c}(Q_{T},Q)} \nonumber \\
&&\times(f_{a/p_{1}}\otimes
\hat{\mathbf{C}}_{ca})(x_{1},Q_{T})(f_{b/p_{2}}\otimes
\hat{\mathbf{C}}_{\bar{c}b})(x_{2},Q_{T}).
\end{eqnarray}
Here $\mu=Q_{T}$ is set to minimize large factor in
$\hat{\mathbf{C}}^{(1)}$. Taking the $Q_{T}^{2}$ integral of
Eq.(\ref{eqt}) and Eq.(\ref{qt}) from $0$ to $Q_{T}^{2}$ and then
expanding them at $\mathcal{O}(\alpha_{s})$, we find
${\mathbf{C}}^{(1)}(z,\frac{b_{0}}{b})=\hat{\mathbf{C}}^{(1)}(z,Q_T)$,
thus we will use Eq.(\ref{m1})-(\ref{m2}) to calculate
${\mathbf{C}}^{(1)}$. However, it should be emphasized that this
formula is valid only with
$\{\mathbf{A}^{(1)},\mathbf{B}^{(1)},\hat{\mathbf{C}}^{(1)}\}$,
and the two formulas must be equal at this order, which suggests a
way to adjust the parameters in $b^*$ prescription \cite{abc}.
Previously, the authors (DDT) of \cite{ddt1} have derived a
similar formula, which is corresponding to our result at
leading-logarithmic-order (LLO). Later, the extended DDT formula
in the CSS frame with $\{A^{(2)},B^{(2)},C^{(1)}\}$ was suggested
in \cite{ddt2}, whose coefficients $\{\tilde{A},\tilde{B},C\}$ at
$\mathcal{O}(\alpha_{s})$ are just our
$\{\mathbf{A}^{(1)},\mathbf{B}^{(1)},\hat{\mathbf{C}}^{(1)}\}$.
\begin{figure}[t!]
\includegraphics{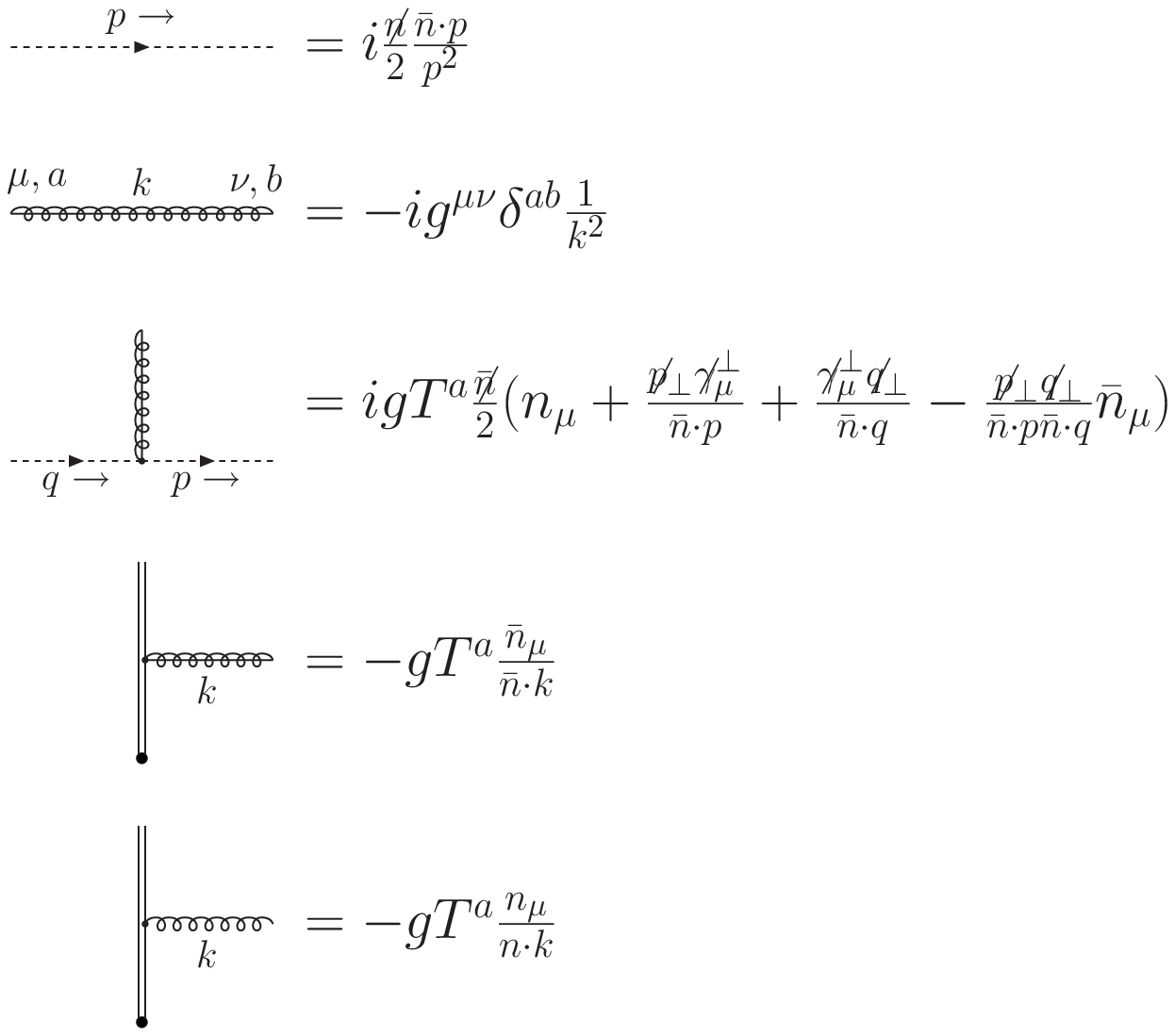}
\caption{\label{feyn} Feynman rules for $n$-direction collinear
particles in Feynman gauge $\alpha=1$ \cite{bsr}. Here the
direction of gluon momentum in the Wilson line is along the
collinear particle. The rules for soft fields and collinear gluon
are the same as that for QCD, and $n\leftrightarrow\bar{n}$ for
$\bar{n}$-direction collinear particles.}
\end{figure}
The real radiative contribution to the differential cross section
at the scale $Q_T^2$ is
\begin{eqnarray}
\nonumber
\frac{d\sigma_{r}^{\mathrm{SCET}}(\mu)}{dQ^{2}dydq_{T}^{2}}&=&
\frac{1}{\Gamma(1-\epsilon)}(\frac{4\pi}{q_T^2})^\epsilon
\int\frac{du}{u}\delta \bigg (\frac{1}{u}(u-u_{min})(u-u_{max}) \bigg )\nonumber\\
&&\times\frac{|\mathcal{M}^{(0)}_{ab}(p_1,p_2,k,y)|^{2}}{8s(2\pi)^{2}}.
\end{eqnarray}
Here $\mathcal{M}^{(0)}_{ab}(p_1,p_2,k,y)$ denotes the
corresponding matrix element, and the usual invariants are defined
as
\begin{eqnarray}
\nonumber s&=&(p_1+p_2)^{2}, \qquad\ t=(p_1-k)^{2},\nonumber\\
u&=&(p_2-k)^{2}, \qquad\ z=\frac{Q^2}{s},\nonumber\
\end{eqnarray}
and the two roots of the equation $(p_1+p_2-q)^{2}=0$ are
$$u_{min}=Q^2\frac{z-1-\sqrt{(1-z)^2-4zq_T^2/Q^2}}{2z},$$
$$u_{max}=Q^2\frac{z-1+\sqrt{(1-z)^2-4zq_T^2/Q^2}}{2z}.$$

\begin{figure}[t!]
\includegraphics{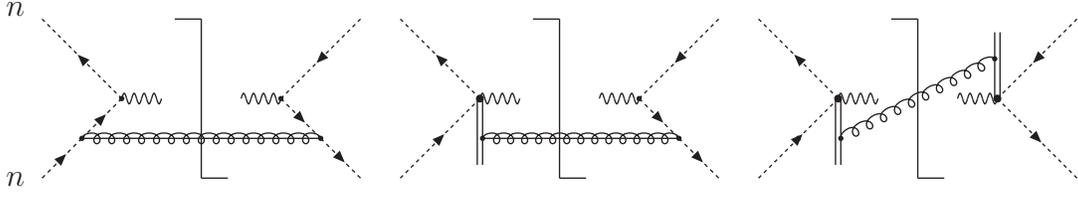}
\caption{\label{c} Cut diagrams for the emission of gluon in
$\mathrm{SCET_{II}}$, and the mirror diagrams are not shown}
\end{figure}

The expression $|\mathcal{M}^{(0)}_{ab}(p_1,p_2,k,y)|^{2}$ in SCET
can be written down by cut diagrams, and the Feynman rules we need
are shown in Fig.\ref{feyn}. For example, in Fig.\ref{c}, the real
contribution of the collinear gluon in the $n$ direction is
\begin{eqnarray}
|\mathcal{M}^{c1(0)}_{qq}(p_1,p_2,k,y)|^{2} &=&
-4\pi\alpha_{s}\mu^{2\epsilon} \frac{1}{36} C_F \bigg
\{\mathrm{Tr} \bigg [s
\gamma_{\mu}\frac{\not{n}}{2}\frac{\bar{n}\cdot
(p_1-k)}{(p_1-k)^2}\frac{\not{\bar{n}}}{2} \\
&& \times \bigg (n_\nu+
\frac{(\not{p_1}-\not{k})_{\perp}\not{\gamma}_\nu^{\perp}}{\bar{n}\cdot
(p_1-k)}+\frac{\not{\gamma}_\nu^{\perp}\not{p_1}_{\perp}}{\bar{n}\cdot
p_1}-\frac{(\not{p_1}-\not{k})_{\perp}\not{p_1}_{\perp}}{\bar{n}\cdot
(p_1-k)\bar{n}\cdot p_1}\bar{n}_\nu \bigg )\frac{\not{n}}{2} \nonumber\\
&&\times \frac{\not{\bar{n}}}{2} \bigg
(n^\nu+\frac{\not{\gamma}_{\perp}^\nu
(\not{p_1}-\not{k})_{\perp}}{\bar{n}\cdot
p_1-k}+\frac{\not{p_1}_{\perp}\not{\gamma}_{\perp}^\nu}{\bar{n}\cdot
p_1}-\frac{(\not{p_1}-\not{k})_{\perp}\not{p_1}_{\perp}}{\bar{n}\cdot
(p_1-k)\bar{n}\cdot
p_1}\bar{n}^\nu \bigg )\frac{\not{n}}{2} \nonumber\\
&&\times \frac{\bar{n}\cdot
(p_1-k)}{(p_1-k)^2}\gamma^{\mu}\frac{\not{\bar{n}}}{2} \bigg
]+2\mathrm{Tr} \bigg [s\gamma_{\mu}\frac{\bar{n}_\nu}{\bar{n}\cdot
k}\frac{\not{n}}{2}\frac{\not{\bar{n}}}{2}n^\nu\frac{\not{n}}{2}\frac{\bar{n}\cdot
(p_1-k)}{(p_1-k)^2}\gamma^{\mu}\frac{\not{n}}{2} \bigg ] \bigg\}\nonumber\\
&=& -4\pi\alpha_{s}\mu^{2\epsilon} \frac{1}{36} \mathrm{Tr}[
s\gamma_{\mu}\frac{\not{n}}{2}\gamma^{\mu}\frac{\not{\bar{n}}}{2}]
2C_F \bigg \{(1-\epsilon) \bigg
[2\frac{(p_1-k)_{\perp}\not{p_1}_{\perp}}{\bar{n}\cdot
(p_1-k)\bar{n}\cdot p_1}\nonumber\\
&&-\frac{(p_1-k)_{\perp}^2}{(\bar{n}\cdot
(p_1-k))^2}-\frac{p_{1\perp}^2}{(\bar{n}\cdot p_1)^2} \bigg ]
\bigg [\frac{\bar{n}\cdot(p_1-k)}{(p_1-k)^2} \bigg
]^{2}+2\frac{\bar{n}\cdot (p_1-k)}{\bar{n}\cdot k (p_1-k)^2} \bigg
\}.\nonumber
\end{eqnarray}
If we drop the common factor
\begin{eqnarray}\mathcal{M} \equiv 4\pi\alpha_{s}\mu^{2\epsilon}
\frac{1}{36}\mathrm{Tr}[Q^{2}\gamma_{\mu}
\frac{\not{n}}{2}\gamma^{\mu}\frac{\not{\bar{n}}}{2}],
\end{eqnarray}
and use the following parametrization for momenta $p_1,p_2,k$,
\begin{eqnarray}
\nonumber p_1 &=& (0,P^-,0), \qquad\ p_2=(P^+,0,0),\nonumber \\
k &=& (\frac{q_T^2}{(1-z_1)P^-},(1-z_1)P^-,-q_T), \\
s &=& P^+P^-, \qquad\ t=\frac{-1}{1-z_1}q_T^2,\nonumber \\
u &=& -(1-z_1)s, \qquad\ tu=sq_T^2,
\end{eqnarray}
the above expression can be simplified to
\begin{eqnarray}
|\mathcal{M}^{c1(0)}_{qq}(p_1,p_2,k,y)|^{2}\rightarrow
\frac{2C_F[(1-\epsilon)(1-z_1)^2+2z_1]}{zq_{T}^{2}}.
\end{eqnarray} The contribution of $\bar{n}$ collinear gluon is
given by $n \leftrightarrow \bar{n}$ and $z_1 \leftrightarrow
z_2$,
\begin{eqnarray}|\mathcal{M}^{c2(0)}_{qq}(p_1,p_2,k,y)|^{2}\rightarrow
\frac{2C_F[(1-\epsilon)(1-z_2)^2+2z_2]}{zq_{T}^{2}},\end{eqnarray}
and the soft gluon contribution is
\begin{eqnarray}|\mathcal{M}^{s(0)}_{qq}(p_1,p_2,k,y)|^{2}
\rightarrow\frac{4C_F}{zq_{T}^{2}}.\end{eqnarray}

The condition that the emitted gluon is to be collinear is
$(1-z_i)P^- \gg q_T$ or $z_i \rightarrow z$ for $i=1,2$, so only
half of the phase space is covered, i.e., $u=u_{min}$ for
$u=-(1-z_i)s$, under which we could safely make the substitution
$z_i \rightarrow z$. The soft gluon is guaranteed by $(1-z_i)P^-
\sim q_T$ or $z \rightarrow 1$. Note that
$|\mathcal{M}^{ci(0)}|^2\rightarrow |\mathcal{M}^{s(0)}|^2$ when
$z_i\rightarrow z\rightarrow 1$, which results in\footnote{We have
divided soft contribution into two parts.}
\begin{eqnarray}
\nonumber |\mathcal{M}^{i(0)}_{qq}(p_1,p_2,k,y)|^{2} &=&
|\mathcal{M}^{ci(0)}_{qq}(p_1,p_2,k,y)|^{2}_{1-z_i\gg \frac{q_T}{P^-}} \nonumber \\
&& +|\mathcal{M}^{s(0)}_{qq}(p_1,p_2,k,y)|^{2}_{1-z_i\sim
\frac{q_T}{P^-}} \nonumber \\
&\rightarrow& \frac{2C_F[(1-\epsilon)(1-z)^{2}+2z]}{zq_{T}^{2}} \nonumber \\
& \equiv & \frac{2(1-z)P_{qq}(z,\epsilon)}{zq_{T}^{2}},
\end{eqnarray}
similarly,
$$|\mathcal{M}^{i(0)}_{ab}(p_1,p_2,k,y)|^{2}\rightarrow \frac{2(1-z)P_{ab}(z,\epsilon)}{zq_{T}^{2}},$$
where
\begin{eqnarray}
P_{qq}(z,\epsilon)&=& C_F\bigg[\frac{1+z^2}{1-z}-\epsilon (1-z)\bigg],\nonumber \\
P_{gq}(z,\epsilon)&=& C_F\bigg[\frac{1+(1-z)^2}{z}-\epsilon z\bigg],\nonumber \\
P_{gg}(z,\epsilon)&=& 2C_A\bigg[\frac{1-z}{z}+\frac{z}{1-z}+z(1-z)\bigg], \\
P_{qg}(z,\epsilon)&=&
T_R\bigg[1-\frac{2z(1-z)}{1-\epsilon}\bigg].\nonumber
\end{eqnarray}
These are the corresponding results of \cite{pt}.

Next we match the $Q_T$ integral Eq.(\ref{m1}) onto PDFs after
making Mellin transformation,
\begin{eqnarray}
\nonumber && \Sigma_{q\bar{q}}(n) \equiv
\frac{\alpha_s}{2\pi}\Sigma_{q\bar{q}}^{(1)}(n)
=\int_{0}^{1-\frac{2q_T}{Q}}dz z^{n}\frac{1}{\sigma^{(0)}}
\frac{d\sigma_{r}^{\mathrm{SCET}}(\mu)}{dQ^{2}dydq_{T}^{2}}, \nonumber \\
&& \Sigma_{q\bar{q}}^{(1)}(n) = \frac{1}{\Gamma(1-\epsilon)}
\frac{1}{q_T^2}(\frac{\mu^2 e^{\gamma_{E}}}{q_T^2})^\epsilon I_n, \\
\label{qti} && I_n = \int_{0}^{1-\frac{2q_T}{Q}}dz
z^{n}\frac{2(1-z)P_{qq}(z,\epsilon)}{\sqrt{(1-z)^2-4z\frac{q_T^2}{Q^2}}},
\end{eqnarray}
where the upper limit $z=1-2q_T/Q$ is to make the integrand
meaningful, and the integral (\ref{qti}) can be evaluated easily
if we retain only the singular terms as $q_T^2\rightarrow 0$,
\begin{eqnarray}
I_n &\rightarrow& \int_{0}^{1-\frac{2q_T}{Q}}dz
\frac{2(1-z)P_{qq}(z,\epsilon)}
{\sqrt{(1-z)^2-4z\frac{q_T^2}{Q^2}}}+\int_{0}^{1}dz(z^n-1)2P_{qq}(z,\epsilon) \nonumber\\
&\rightarrow&
2C_F\log\frac{Q^{2}}{q_T^2}-3C_F+2\gamma_{qq}(n)+2\epsilon
\gamma_{qq}^{\epsilon}(n),
\end{eqnarray}
where $\gamma_{qq}(n)$ and $\gamma_{qq}^{\epsilon}(n)$ are the
moment of regularized splitting function and of
$P_{qq}^{\epsilon}(z)$. So
\begin{eqnarray}
\nonumber \int_0^{Q_T^2}dq_T^2\Sigma_{q\bar{q}}^{(1)}(n)
&=&\frac{1}{\Gamma(1-\epsilon)}
\int_{0}^{Q_T^2}\frac{dq_T^2}{q_T^2}(\frac{\mu^2
e^{\gamma_{E}}}{q_T^2})^\epsilon
\left[2C_F\log\frac{Q^{2}}{q_T^2}-3C_F+2\gamma_{qq}(n)+2\epsilon
\gamma_{qq}^{\epsilon}(n)\right] \nonumber \\
&=&\frac{2C_{F}}{\epsilon^2}+\frac{1}{\epsilon}(3C_{F}
+2C_{F}\log\frac{\mu^{2}}{Q^{2}})-\frac{2\gamma_{qq}(n)}{\epsilon}
-2\gamma_{qq}^{\epsilon}(n)-\frac{C_{F}\pi^2}{6} \\
&& + C_{F}\log^2\frac{\mu^{2}}{Q_T^{2}}
+3C_{F}\log\frac{\mu^{2}}{Q_T^{2}}
-2\gamma_{qq}(n)\log\frac{\mu^{2}}{Q_T^{2}}
-2C_{F}\log\frac{Q^{2}}{Q_T^{2}}\log\frac{\mu^{2}}{Q_T^{2}}.\nonumber
\end{eqnarray}

The virtual correction comes from the UV renormalization constant,
\begin{eqnarray}
\nonumber 2 \delta Z_V=\frac{\alpha_s}{2\pi}\left[\frac{2C_{F}}
{\epsilon^2}+\frac{1}{\epsilon}(3C_{F}
+2C_{F}\log\frac{\mu^{2}}{Q^{2}})\right],
\end{eqnarray}
which cancels the first two $\epsilon$-poles. The remaining is
cancelled by the renormalization of PDF, i.e.,
\begin{eqnarray}
f_q(n)&=& f_{\bar{q}}(n) = 1+\frac{\alpha_s}{4\pi}
\left[2\gamma_{qq}(n)(\frac{1}{\epsilon_{UV}}
-\frac{1}{\epsilon_{IR}})\right],
\end{eqnarray}
where $\mu=\mu_F$ is implied. Finally, we get
\begin{eqnarray}
\hat{\mathbf{C}}_{qq}^{(1)}(n,Q_T)&=&-2\gamma_{qq}^{\epsilon}(n)-\frac{C_{F}\pi^2}{6},\\
\mathbf{C}^{(1)}_{qq}(z,\frac{b_{0}}{b})&=&-2P_{qq}^{\epsilon}(z)-\frac{C_{F}\pi^2}{6}\delta(1-z),
\end{eqnarray}
Similarly,
\begin{eqnarray}
\mathbf{C}^{(1)}_{ab}(z,\frac{b_{0}}{b})=-2P_{ab}^{\epsilon}(z)-\frac{C_{a}\pi^2}{6}\delta_{ab}\delta(1-z).
\end{eqnarray}

\end{document}